\documentclass{article}
\usepackage[utf8]{inputenc}
\usepackage{graphicx} 
\usepackage{amsmath}
\usepackage{amssymb}
\usepackage[margin=1in]{geometry}
\usepackage{appendix}
\usepackage[numbers]{natbib}
\usepackage{subcaption}

\title{On the practicalities of producing a nuclear weapon using high-assay low-enriched uranium}
\author{P. Cosgrove, N. Read}
\date{August 2024}

\begin{document}

\maketitle

It was recently argued by Kemp et al.~that HALEU (high-assay low-enriched uranium, or uranium enriched up to 19.75\%) can conceivably be used to produce a nuclear weapon and on this basis civilian enrichment limits should be lowered to 10\% or 12\% \citep{Kemp2024}. We find their argument unconvincing in several respects.

The question of maximum allowable enrichment should be resolved by a risk-benefit appraisal: finding an optimal point that allows useful peaceful applications of nuclear technology while reducing proliferation risks to an acceptable level. We believe that the impact of reducing maximum enrichment on nuclear technology is real and meaningful, but was almost completely dismissed by Kemp et al. Before addressing this issue, we shall discuss the article's main focus which was the risks presented by HALEU.

Kemp et al. assert that `the practical limit for weapons lies below the 20\% threshold' and proceed to estimate the yield of such a weapon. They report using the Bethe-Feynman formula. The applicability of this formula to the \textless 20\% enrichment case under consideration is dubious. The formula is simple, designed to provide a qualitative indication of a weapon's performance without requiring expensive time-dependent neutronics and hydrodynamics calculations. The common formula is~\citep{Lestone2022}:
\begin{equation*}
    Y \approx f M R_0^2\alpha^2_0\delta\;\mathrm{,}
\end{equation*}
where $Y$ is the yield, the subscript $0$ denotes quantities evaluated at the point of maximum compression of the weapon, $f$ is a dimensionless scaling factor which does not have a commonly stated value but sister formulae tend to have $f\sim0.3$ \citep{Lestone2022}, $M$ is the mass of fissile material, $R_0$ is the radius of the fissile material at maximum compression, $\alpha_0$ is the exponential time constant for neutron multiplication at maximum compression, and $\delta$ is the relative change in radius between $R_0$ and the radius at `second criticality', $R_2$:
\begin{equation*}
    \delta = \frac{R_2}{R_0} - 1\;\mathrm{.}
\end{equation*}
Second criticality is the point at which the weapon has begun to disassemble to the extent that the system becomes sub-critical again, i.e., net neutron multiplication ceases and the neutron population reaches its maximum. The original derivation of this formula appears to be classified \citep{Rosen2021}, but there are numerous derivations of sister formulae~\citep{Reed2007,Reed2016,Lestone2022} and an alternative derivation which does not appear to reproduce commonly accepted values of $f$\citep{Glasstone1972}.

Using HALEU implies several significant changes to the inputs of the formula compared to more typical systems. Due to the large quantities of $^{238}$U, increased neutron capture and inelastic scattering will reduce the value of $\alpha_0$. Given the achievable compressions will not be larger than for enriched material, the mass required will be larger given less favourable neutronic properties. This also implies that the compressed and critical radii will be larger. 

There is conflicting information on the approximations which are used in deriving Bethe-Feynman. Lestone et al.~\cite{Lestone2022} mention that an ansatz is used for the variation in $\alpha$ with the expanding radius of the problem, while Glasstone and Redman \cite{Glasstone1972} say that $\alpha$ is presumed to be essentially constant at $\alpha_0$. However, it appears that the derivations do not account for variations in $\alpha$ due to the evolution of the initiating shock wave (separate from variations driven by releasing fission energy). This approximation is likely to be invalid in the case of HALEU.

$\alpha$ can be understood as the net neutron production rate per neutron per unit time, with common units of $1/\mu\mathrm{s}$. Starting from an initial neutron at time $t=0$, the total fission energy released in a system at some future time is given by
\begin{equation*}
    E(t) = E_\mathrm{f}\mathrm{e}^{\int^t_0\alpha(t)\mathrm{d}t}\;\mathrm{,}
\end{equation*}
where $E_\mathrm{f} \sim 200 \mathrm{MeV}$ is the energy released per fission. From this equation, when $\int^t_0\alpha(t)\mathrm{d}t \sim 53$, a kiloton of energy will have been produced. For more standard weapons material, the value of $\alpha_0$ tends to be on the order of 100 $\mu \mathrm{s}^{-1}$ or greater~\citep{Mark1993}. This can be verified with approximate neutron cross sections available in the literature for weapons material or by more specialised calculations of $\alpha$ which are also present in the literature~\citep{Cullen2003,Zoia2014}. Such values of $\alpha_0$ imply that, for an approximately constant $\alpha(t) = \alpha_0$, a nuclear explosion will be produced in less than a microsecond if neutrons are introduced to the system at peak compression. The speed at which an explosion is achieved may go some way to justifying approximating $\alpha_0$ as constant in some Bethe-Feynman derivations \cite{Glasstone1972}.

HALEU enriched to 19.75\%, even when subjected to the upper bounds of realistic compressions, cannot achieve such large values of $\alpha_0$. For approximate calculations, $\alpha_0$ depends sensitively on the choice of average neutron energy in the system due to $^{238}$U's threshold fission reaction. However, using a modified Monte Carlo neutron transport code we did not observe any value greater than about 25 $\mu \mathrm{s}^{-1}$ given generous compression and a heavy tamper. Less generous compression and tamping (which may be more realistic for a necessarily larger device than usual) gave values closer to 5 $\mu \mathrm{s}^{-1}$. These values imply a dramatic lengthening of the neutronic timescales. These timescales may even become comparable to those of the compression and decompression which \cite{Mark1993} suggests to be on the order of 10 $\mu\mathrm{s}$.

Consider how $\alpha$ will vary in a system without introducing neutrons: the shock compression will cause $\alpha$ to increase until the shock reaches the centre before passing through, causing $\alpha$ to fall again. 
For a system where $\alpha_0$ is large, introducing neutrons at peak compression will result in many fission neutrons being produced quickly, before the system is explosively disassembled -- this should result in a meaningfully large yield.
However, if $\alpha_0$ is small, it is conceivable that $\alpha$ may have decreased from $\alpha_0$ before a sufficient number of fissions have occurred. This means that $\alpha_0$ is not representative of the multiplication rate for this system. One could, in principle, carefully time the introduction of neutrons to allow the population to grow such that large energy production only occurs at peak compression. However, this would require precise understanding of the hydrodynamics and careful control of the neutron introduction time. In the most extreme case, for small but positive values of $\alpha_0$, it is conceivable that the weapon will have decompressed before any significant multiplication occurs, regardless of how early neutrons are introduced.

\begin{figure}[h!]
    \centering
    \begin{subfigure}[b]{0.32\textwidth}
        \centering
        \includegraphics[width=0.9\textwidth]{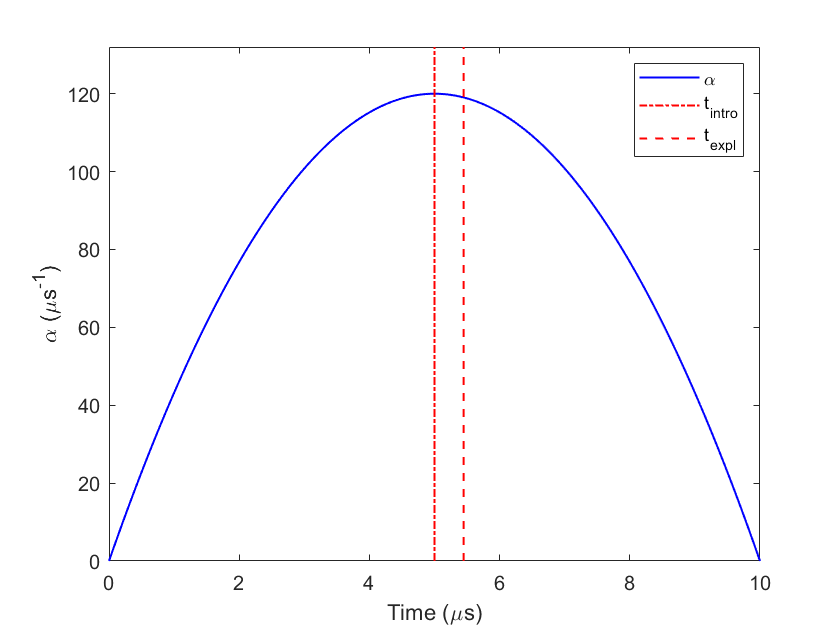}
        \caption{$\alpha_0 = 120\mu\mathrm{s}^{-1}$}
    \end{subfigure}%
    ~ 
    \begin{subfigure}[b]{0.32\textwidth}
        \centering
        \includegraphics[width=0.9\textwidth]{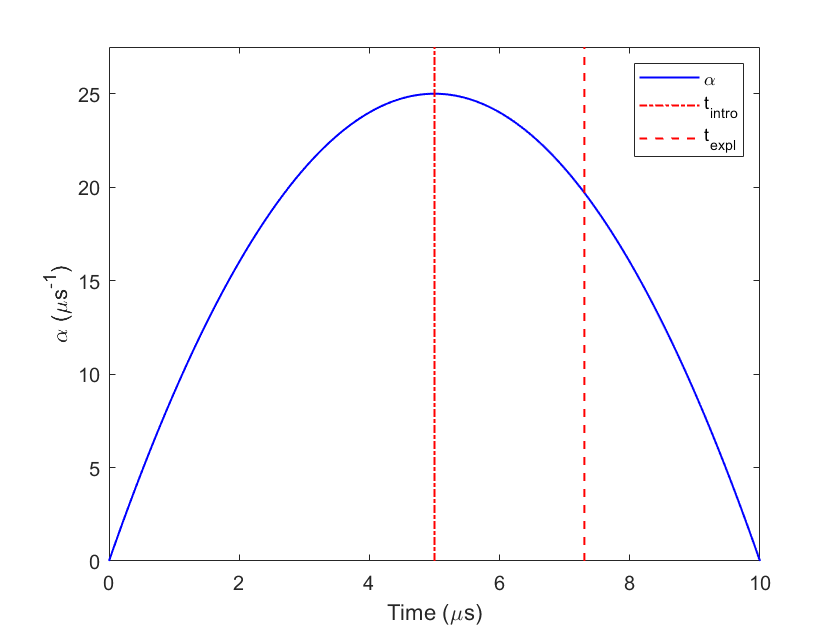}
        \caption{$\alpha_0 = 25\mu\mathrm{s}^{-1}$}
    \end{subfigure}
    ~ 
    \begin{subfigure}[b]{0.32\textwidth}
        \centering
        \includegraphics[width=0.9\textwidth]{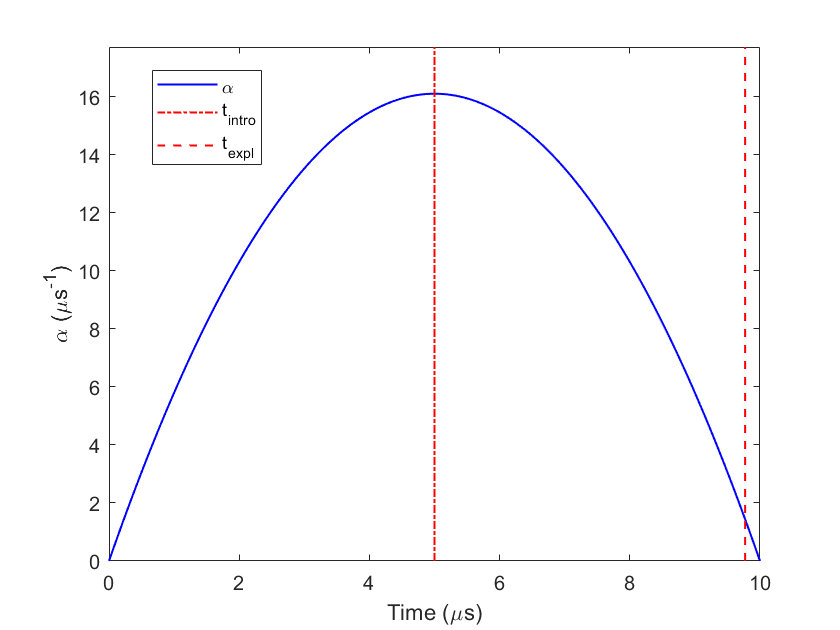}
        \caption{$\alpha_0 = 16\mu\mathrm{s}^{-1}$}
    \end{subfigure}
    \caption{Illustrating the time to explosion for different values of $\alpha_0$, assuming $\alpha$ varies quadratically in time, that neutrons are introduced at peak compression, and neglecting the drop in $\alpha$ due to explosive disassembly. $t_\mathrm{intro}$ is the time at which neutrons are introduced while $t_\mathrm{expl}$ is the time at which $\int^{t_\mathrm{expl}}_{t_\mathrm{intro}}\alpha(t)\mathrm{d}t = 53$}
    \label{fig:alpha}
\end{figure}

This is illustrated in Fig.~\ref{fig:alpha}, where the time to explosion is estimated for three different values of $\alpha_0$, assuming a quadratic variation in $\alpha$ with time, that neutrons are introduced at peak compression, and neglecting the rapid decrease in $\alpha$ due to explosive forces. For such a system, a yield could in theory be achieved provided $\alpha_0 = 16\mu\mathrm{s}^{-1}$, but the rapid drop in $\alpha$ in real systems~\citep{Glasstone1972} implies that either a higher value of $\alpha$ or more precise control over the time of neutron introduction would be necessary.

If one were to use the Bethe-Feynman formula without noting its limitations, it would produce a substantial yield for HALEU -- the formula is quadratic in the radius of the necessarily larger weapon and linear in the mass. However, it would also produce a substantial yield for more easily-attainable 6\% enriched uranium (LEU) -- this is about the minimum enrichment for uranium to be able to sustain a fast neutron chain reaction, given a sufficiently large mass. Using Bethe-Feynman, a strong compression of 10t of LEU would produce a roughly 50kt explosion (assuming pre-detonation is overcome~\citep{Kemp2024}\citep{Mark1993}). 

In theory, any material with a finite bare critical mass can be used to make a nuclear explosion, but the common version of the Bethe-Feynman formula does not appear to be a reasonable basis on which to evaluate such a weapon's practicality. Despite \cite{Kemp2024} highlighting Bethe's favourable view of the formula, no convincing reason is given why it is valid to apply it to a completely different regime than the one for which it was derived.

On closer inspection of Kemp et al.'s cited testimony of J. Carson Mark~\citep{Mark1984}, we believe that he takes a similarly pragmatic view, less categorical than that implied by the original piece. To quote him more fully, he says `it is possible on paper to imagine' producing an explosion down to 10\% enrichment but cautions that `the penalties, however, are quite tremendous' in terms of weapons design. He also highlights that `HEU is really a whole area of different materials' and warns against `blind black and white categorisation'. 

Mark's testimony appropriately concludes by stating that `making a bomb is far from as easy as some people have sometimes said'. Technical difficulty is presently considered to be a barrier in proliferating, e.g., requiring a reprocessing facility to extract plutonium from spent nuclear fuel. Given that a HALEU implosion-type weapon has never -- to our knowledge -- been tested by weapons labs, it seems reasonable to suggest that this, too, is a difficult path towards producing a weapon. It is also notable that Mark endorsed considering the use of 50\% enriched fuel in research reactors, despite the hearing's proposal being for 20\% \citep{Mark1984}.

Quantifying the difficulty of making a HALEU weapon is nearly impossible without classified information. We do not object to the calls of Kemp et al.~for more clarity from those who do possess such knowledge regarding the true difficulty of making such a device so that it can be placed in context with other non-proliferation risks. They call for a new study, which is reasonable, but there is an unreasonable presumption that the outcome ``should be... a lower enrichment limit". The selective use of quotations from J. Carson Mark and others, combined with a dubious application of the Bethe-Feynam formula, is not enough to form such an expectation.

Context is vital here, since we do accept some level of risk of proliferation of many kinds in exchange for the benefits of nuclear technology. A HALEU weapon may be `possible to imagine on paper', but we note that reactor-grade plutonium is not just possible on paper, it has been demonstrated to work \cite{doe_facts_additional_1994}, although undoubtedly with many technical challenges. Under the current safeguards regime, the risks presented by this plutonium are generally deemed acceptable, especially given that the cost of removing this risk would likely be the discontinuation of uranium-based fission energy in general.

What, then, would the cost be of removing the risk posed by HALEU by adopting a 10\% or 12\% limit, as Kemp et al.~suggest?

Kemp et al.~correctly point out that higher enrichments are desirable for relaxing constraints in the reactor design process, and for some designs 20\% is required simply to sustain a chain reaction. However, later in the article  it is asserted that the costs of limiting enrichment to 10\% or 12\% would ``allow many reactor designs to move forward with only modest economic consequences''. This single sentence is all that is offered to address the wide-ranging and variable consequences for reactor technology of limiting enrichment. Its rather vague notion that many reactors can `move forward', has the unstated corollary that other reactors designs will have to stop. Indeed this is the case. For those designs that can remain possible, the claim of `modest costs' is substantiated by an entirely unconvincing reference. The document in question is the proceedings arising from a workshop focused on a very new and niche segment of the nuclear energy domain: fission batteries, also known as `microreactors'. These systems typically have power outputs in the range of 1-10 megawatts of electrical power, as compared to up to 300 megawatts for small modular reactors or up to 1600 megawatts for large conventional power plants. The workshop was ``a first assessment of the markets, economics, and business models for [microreactors]. This is a work in progress''. Enrichment is discussed very little in these proceedings. The only conceivable source of the claim that 10\% enrichment imposes modest costs comes from a high-level analysis of microreactor economics which finds that \textbf{for microreactors} it is preferable to have designs that have lower fuel costs by staying roughly below 10\% enrichment, since there is little economic incentive to pursue long-lifetime systems that would require higher enrichment. This finding has no applicability outside of the specific microreactor context in which it was derived.

The costs of imposing a lower enrichment limit on reactors are far more wide ranging and complex, requiring in-depth discussion. They cannot be dismissed in one sentence. Two aspects will be covered briefly here but the discussion will be far from exhaustive. First, some reactors are indeed rather inefficient with their neutrons, as Kemp et al. state. In the case of very small systems, many neutrons leave the core and are absorbed in the shielding, rather than producing more fission reactions. Higher enrichments are therefore required if we wish to produce the very smallest reactor sizes. Another popular design choice is TRISO fuel, a particulate fuel form composed of sub-millimetre fuel spheres surrounded by protective layers. This fuel has useful properties, allowing operation at high temperatures that lead to high efficiency of electricity production and the possibility of directly decarbonising carbon-intensive industrial processes. They also have exceptional safety performance, being capable of tolerating the extreme temperatures arising in various accident scenarios. Unfortunately, the high surface area of fuel leads to more neutrons being absorbed by the non-fissile $^{238}$U isotope. Often such reactors also have long, narrow reactor cores that have a higher surface area for heat removal, but this also leads to further neutron losses. These systems are able to passively shut down and remove all heat with no requirement for electrical power. The higher enrichments (and associated fuel costs) required to compensate for the neutron losses are acceptable, since it is presumed that the value of high temperature heat and highly reliable safety systems will be an overall economic gain. Within this category of reactors, the larger systems should be able to operate with less than 10\% enrichment, but the smaller ones will be unviable.

The second family of reactors affected, and more severely so, is fast-spectrum systems (`fast reactors'), such as the Natrium reactor mentioned by Kemp et al. These can not be said to be inefficient with their neutrons. Indeed, they produce an abundance of neutrons that can be used for other purposes. Historically, the main use of interest for these excess neutrons was to `breed' more fissile material, to maximise the amount of energy extracted from natural uranium resources. Due to the surprising abundance of uranium, the motivation to pursue these systems has reduced in the last few decades. However, the possibility of transmuting long-lived radioisotopes into shorter-lived ones still receives attention for its possible benefits for waste disposal. They also generally have high operating temperatures, with the benefits described above. Many also have the same attractive passive shutdown and decay heat removal features as TRISO-based systems. In the future, if nuclear is used at larger scales, the resource utilisation motivations may return. Despite their neutronic efficiency, for reasons of reactor physics, such reactors do simply require higher concentrations of fissile material to sustain a chain reaction. If this is provided by uranium, enrichments higher than 12\% are necessary. By limiting enrichment, fast reactors become either impossible, or must rely on their traditionally-assumed choice of fissile material: plutonium. It is questionable whether this would be a net benefit from a non-proliferation perspective.

The prior points suggest that Kemp et al.~take a markedly one-sided view of the costs and benefits of HALEU. The risks of a HALEU device have been exaggerated, the opinions of experts have been only partially provided, and what these experts view as an extremely difficult device is dismissed by Kemp et al.~as `not without its challenges'. Meanwhile the negative impacts on nuclear energy have essentially been ignored, even where these technologies can meaningfully contribute to decarbonising electricity and industrial processes at scale. The requests for clarity and review from Kemp et al.~are reasonable, but it is far from clear that the outcomes of such a review would be a recommendation to reduce the enrichment limit.

\bibliographystyle{unsrt}
\bibliography{main}

\begin{thebibliography}{10}

\bibitem{Kemp2024}
R.~Scott Kemp, Edwin~S. Lyman, Mark~R. Deinert, Richard~L. Garwin, and Frank~N. von Hippel.
\newblock The weapons potential of high-assay low-enriched uranium.
\newblock {\em Science}, 384(6700):1071--1073, 2024.

\bibitem{Lestone2022}
J.P. Lestone, M.D. Rosen, and P.~Adsley.
\newblock {Comparison Between Historic Nuclear Explosion Yield Formulas}.
\newblock {\em Nuclear Technology}, 207(sup1):S352--S355, 2021.

\bibitem{Rosen2021}
M.~Rosen.
\newblock {An Analytic Derivation of the Bethe-Feynman (BF) Formalism}.
\newblock {\em Weapons Review Letters}, 2021.

\bibitem{Reed2007}
B.~Cameron Reed.
\newblock {Arthur Compton’s 1941 Report on explosive fission of U-235: A look at the physics}.
\newblock {\em American Journal of Physics}, 75(12):1065--1072, 12 2007.

\bibitem{Reed2016}
B~Cameron Reed.
\newblock {A physicists guide to The Los Alamos Primer}.
\newblock {\em Physica Scripta}, 91(11):113002, oct 2016.

\bibitem{Glasstone1972}
S.~Glasstone and L.M. Redman.
\newblock An introduction to nuclear weapons.
\newblock Technical Report WASH-1037, U.S. Atomic Energy Commission, Division of Military Applications, June 1972.

\bibitem{Mark1993}
J.~Carson Mark.
\newblock Explosive properties of reactor‐grade plutonium.
\newblock {\em Science \& Global Security}, 4(1):111--128, 1993.

\bibitem{Cullen2003}
D.E. Cullen, C.J. Clouse, R.~Procassini, and R.C. Little.
\newblock {Static and Dynamic Criticality: Are They Different?}
\newblock Technical Report UCRL-TR-201506, Lawrence Livermore National Laboratory, November 2003.

\bibitem{Zoia2014}
Andrea Zoia, Emeric Brun, and Fausto Malvagi.
\newblock {Alpha eigenvalue calculations with Tripoli-4®}.
\newblock {\em Annals of Nuclear Energy}, 63:276--284, 2014.

\bibitem{Mark1984}
{Conversion of Research and Test Reactors to Low-enriched Uranium (LEU) Fuel}, 1984.
\newblock Hearing Before the Subcommittee on Energy Development and Applications and the Subcommittee on Energy Research and Production of the Committee on Science and Technology, U.S. House of Representatives, Ninety-eighth Congress, Second Session.

\bibitem{doe_facts_additional_1994}
U.S.~Department of~Energy.
\newblock Additional {Information} {Concerning} {Underground} {Nuclear} {Weapon} {Test} of {Reactor}-{Grade} {Plutonium}, 1994.

\end{thebibliography}

\end{document}